\documentclass[%
 reprint,
superscriptaddress,
amsmath,amssymb,
aps,
pra,
]{revtex4-1}
\usepackage{caption}
\usepackage{subcaption}
\usepackage{graphicx}
\usepackage{xcolor}
\usepackage{graphicx}
\usepackage{dcolumn}
\usepackage{bm}

\begin{document}

\preprint{APS/123-QED}

\title{Refraction of space-time wave packets: III.~Experiments at oblique incidence}

\author{Murat Yessenov}
\affiliation{CREOL, The College of Optics \& Photonics, University of Central Florida, Orlando, FL 32816, USA}
\author{Alyssa M. Allende Motz}
\affiliation{CREOL, The College of Optics \& Photonics, University of Central Florida, Orlando, FL 32816, USA}
\author{Basanta Bhaduri}
\affiliation{CREOL, The College of Optics \& Photonics, University of Central Florida, Orlando, FL 32816, USA}
\author{Ayman F. Abouraddy}
\thanks{corresponding author: raddy@creol.ucf.edu}
\affiliation{CREOL, The College of Optics \& Photonics, University of Central Florida, Orlando, FL 32816, USA}




\begin{abstract}
The refraction of space-time (ST) wave packets at planar interfaces between non-dispersive, homogeneous, isotropic dielectrics exhibit fascinating phenomena, even at normal incidence. Examples of such refractive phenomena include group-velocity invariance across the interface, anomalous refraction, and group-velocity inversion. Crucial differences emerge at oblique incidence with respect to the results established at normal incidence. For example, the group velocity of the refracted ST wave packet can be tuned simply by changing the angle of incidence. In paper (III) of this sequence, we present experimental verification of the refractive phenomena exhibited by ST wave packets at oblique incidence that were predicted in paper (I). We also examine a proposal for `blind synchronization' whereby identical ST wave packets arrive simultaneously at different receivers without \textit{a priori} knowledge of their locations except that they are all located at the same depth beyond an interface between two media. A first proof-of-principle experimental demonstration of this effect is provided. 
\end{abstract}

\maketitle

\section{Introduction}

Snell's law governs the change in the propagation direction of a monochromatic plane wave incident onto a planar interface between two optical media \cite{SalehBook07}. If $n_{1}$ and $n_{2}$ are the refractive indices of the two media, and $\phi_{1}$ and $\phi_{2}$ are the propagation angles with respect to the normal to the interface for the incident and refracted waves, respectively, then Snell's law dictates that $n_{1}\sin{\phi_{1}}\!=\!n_{2}\sin{\phi_{2}}$. Although this result, strictly speaking, applies to \textit{only} monochromatic plane waves, nevertheless its utility is typically extended in practice to conventional pulsed beams where it can provide an adequate approximation, especially for narrowband paraxial fields in which the spatial and temporal degrees of freedom (DoFs) are uncoupled. Crucially, this entails that the group velocity of the transmitted wave packet depends solely on the local optical properties of the second medium, and is independent of the incident angle. In other words, no `memory' of the incident wave packet is retained as far as the group velocity of the transmitted wave packet is concerned.

In contrast to conventional wave packets, space-time (ST) wave packets \cite{Kondakci16OE,Parker16OE} are a unique class of pulsed optical beams in which the spatial and temporal DoFs are \textit{non}-separable, and are instead inextricably intertwined \cite{Reivelt03arxiv,Kiselev07OS,Turunen10PO,FigueroaBook14}. ST wave packets are endowed with a precise spatio-temporal structure in which each spatial frequency is tightly associated with a single wavelength (or temporal frequency) \cite{Donnelly93ProcRSLA,Saari04PRE,Longhi04OE,Kondakci19OL,Hall21OL}. Uniquely, this spatio-temporal structure determines the group velocity of the ST wave packet \cite{Salo01JOA,Recami03IEEEJSTQE,Valtna07OC,Zamboni08PRA,Kondakci17NP,Efremidis17OL,Wong17ACSP2,Kondakci19NC,Bhaduri19Optica}. Consequently, the rearrangement of the field structure upon refraction at a planar interface leads to a change in the group velocity of the transmitted wave packet that depends on the group velocity of the incident wave packet. Because of this, ST wave packets -- in contrast to conventional wave packets -- retain a `memory' of the incident wave packet, which is the basis for the fascinating refractive phenomena exhibited by ST wave packets at normal and oblique incidence \cite{Bhaduri20NP}.

In paper (I) of this series, we presented a theoretical study of the refraction of ST wave packets at normal and oblique incidence on a planar interface  between two non-dispersive, homogeneous, isotropic dielectrics \cite{Yessenov21RefractionI}. At normal incidence, the above-described memory effect is the basis for group-velocity invariance, anomalous refraction, and group-velocity inversion \cite{Bhaduri20NP}. These phenomena extend to oblique incidence, although the conditions for their realization change with incident angle. Crucially, a new refractive phenomenon emerges at oblique incidence: the group velocity of the transmitted wave packet changes with the incident angle when all else is held fixed. Indeed, the group velocity increases with incident angle upon refraction into a higher-index medium when the incident wave packet is subluminal, and it decreases when the wave packet is superluminal \cite{Bhaduri20NP,Yessenov21RefractionI}.

In paper (II), we provided experimental confirmation of the predicted phenomena at normal incidence \cite{Yessenov21RefractionII}. Here we examine experimentally in detail the refraction of ST wave packets at \textit{oblique} incidence at a planar interface between two non-dispersive, homogeneous, isotropic dielectrics using the interferometric measurement strategy described in paper (II). After a brief overview of the basic law of refraction governing baseband ST wave packets at oblique incidence, we confirm the predicted dependence at oblique incidence of the group velocity for the transmitted wave packet on that of the incident wave packet and on the incidence angle. We then verify the predicted changes that occur at oblique incidence in the conditions required to realize group-velocity invariance and group-velocity inversion. Finally, we examine a proposal that makes use of these unique characteristics to realize \textit{blind synchronization} of multiple receivers via pulses emitted from a single transmitter with no \textit{a priori} knowledge of the receiver positions except that they are situated at the same depth below an interface between two media. Building on our recent observation of isochronous ST wave packets that traverse a planar slab at a fixed group delay independently of the angle of incidence (despite the change in distance traversed) \cite{Motz21arxiv}, we report here a proof-of-principle realization of the blind synchronization scheme.

\begin{figure}[t!]
\centering
\includegraphics[width=8.6cm]{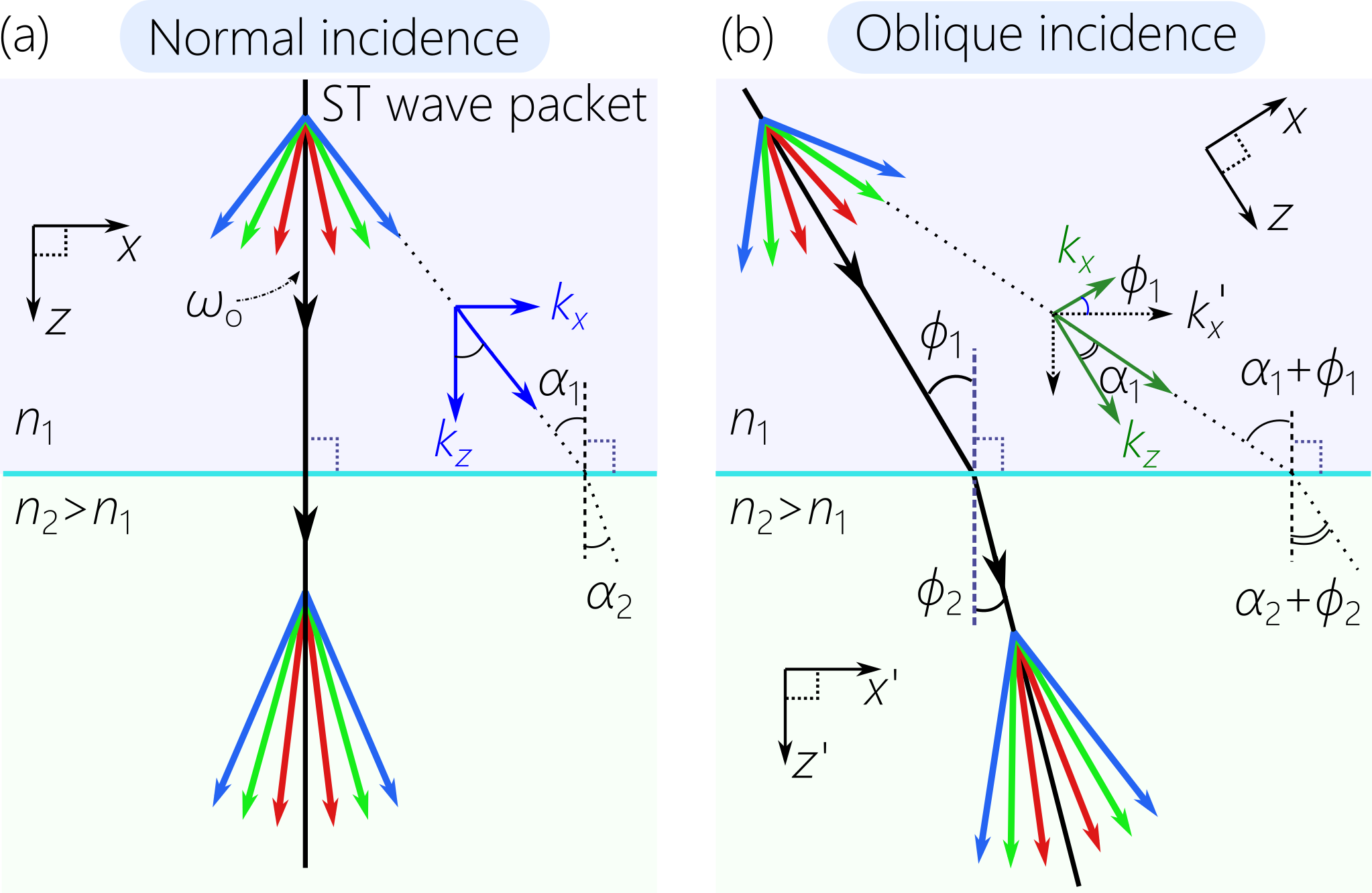}
\caption{(a) Normal and (b) oblique incidence of a ST wave packet on a planar interface.}
\label{Fig:NormalVsObliqueIncidence}
\end{figure}

\begin{figure*}[t!]
\centering
\includegraphics[width=17.4cm]{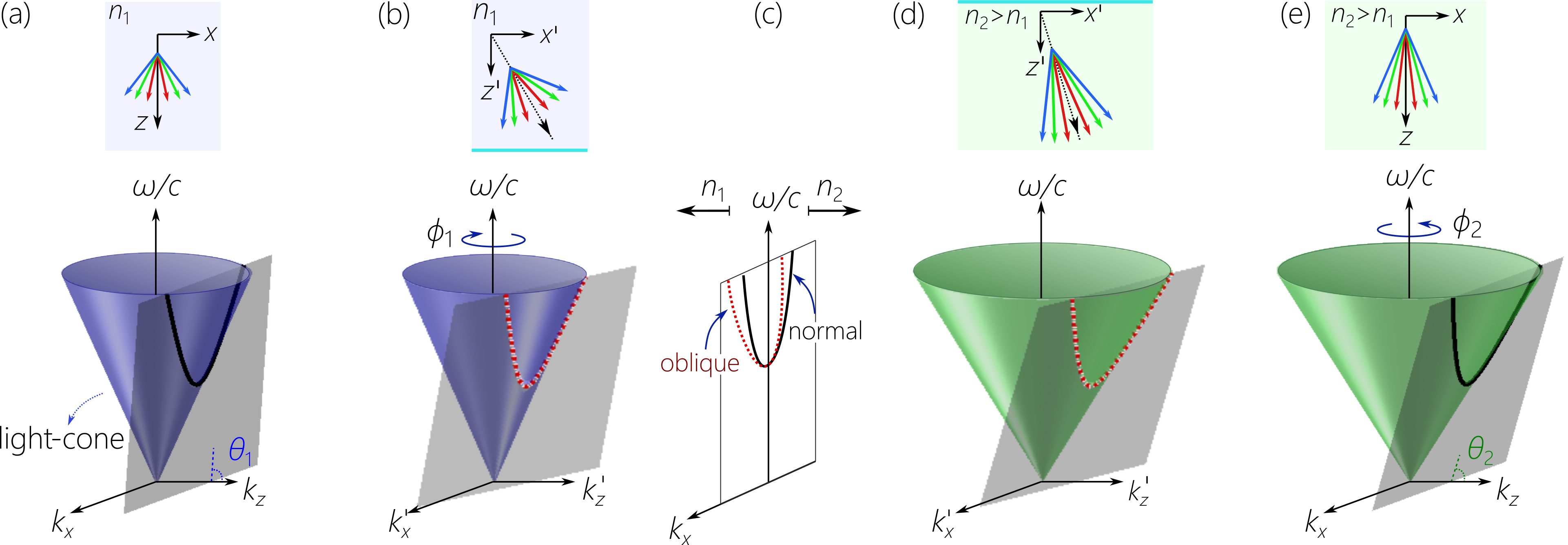}
\caption{(a) Representation of the spectral support domain of a ST wave packet in a medium of refractive index $n_{1}$ in the $(x,z)$ coordinate system that is aligned with the propagation direction of the ST wave packet. The spectrum lies at the intersection of the light-cone with a plane that is parallel to the $k_{x}$-axis and makes an angle $\theta_{1}$ with respect to the $k_{z}$-axis in $(k_{x},k_{z},\tfrac{\omega}{c})$ space. The $z$-axis is aligned with the plane-wave component at $\omega\!=\!\omega_{\mathrm{o}}$ and $k_{x}\!=\!0$. (b) The spectral support domain at oblique incidence represented in the coordinate system $(x',z')$ that is aligned with the interface but \textit{not} with the propagation direction of the wave packet. The spectral support domain is rotated by $\phi_{1}$ (the incident angle) with respect to the $\tfrac{\omega}{c}$-axis in $(k_{x}',k_{z}',\tfrac{\omega}{c})$ space. (c) The spectral projection onto the $(k_{x}',\tfrac{\omega}{c})$-plane from (b) is invariant across the interface at oblique incidence (dotted parabolic curve centered at $k_{x}'\!\neq\!0$). At normal incidence, the spectral projection onto the $(k_{x},\tfrac{\omega}{c})$-plane from (a) is invariant across the interface (solid parabolic curve centered at $k_{x}\!=\!0$). (d) Spectral support domain of the transmitted ST wave packet after oblique incidence. (e) Representation of the spectral support domain of a ST wave packet in a medium of refractive index $n_{2}$. The new spectral tilt angle $\theta_{2}$ is related to $\theta_{1}$ via Eq.~\ref{Eq:ObliqueIncidenceLaw}.}
\label{Fig:Configuration}
\end{figure*}

\section{Law of refraction for space-time wave packets at oblique incidence}

A ST wave packet is a pulsed beam having finite spatial and temporal bandwidths in which each spatial frequency $k_{x}$ is associated with a single temporal frequency $\omega$; here $x$ and $z$ are the transverse and axial coordinates, respectively, and $k_{x}$ and $k_{z}$ are the corresponding components of the wave vector. We identify the propagation direction with the $z$-axis, we assume a spatial spectrum that is symmetric around $k_{x}\!=\!0$, and we assign the temporal frequency $\omega_{\mathrm{o}}$ to the spatial frequency $k_{x}\!=\!0$. The spectral support domain for a `baseband' ST wave packet \cite{Yessenov19PRA} in a non-dispersive medium of index $n$ lies at the intersection of the light-cone $k_{x}^{2}+k_{z}^{2}\!=\!n^{2}(\tfrac{\omega}{c})^{2}$ with a tilted plane $\omega\!=\!\omega_{\mathrm{o}}+(k_{z}-nk_{\mathrm{o}})c\tan{\theta}$ that is parallel to the $k_{x}$-axis and makes an angle $\theta$ (the spectral tilt angle) with respect to the $k_{z}$-axis, where $k_{\mathrm{o}}\!=\!\omega_{\mathrm{o}}/c$ and $c$ is the speed of light in vacuum \cite{Yessenov19OE,Yessenov19PRA,Yessenov19OPN}. This construction results in a propagation-invariant wave packet transported rigidly at a group velocity $\widetilde{v}\!=\!c\tan{\theta}\!=\!c/\widetilde{n}$, where $\widetilde{n}\!=\!\cot{\theta}$ is the group index. In other words, the group velocity is determined by an internal DoF (the spectral tilt angle $\theta$) that is related to the spatio-temporal field structure. Note that the ST wave packet reverts to a plane-wave pulse when $\widetilde{n}\!=\!n$. Within this framework, $\omega_{\mathrm{o}}$ is either the minimum temporal frequency when the wave packet is superluminal ($\widetilde{v}\!>\!c/n$), or the maximum temporal fewquency when the wave packet is subluminal ($\widetilde{v}\!<\!c/n$) \cite{Yessenov19PRA}.

When the ST wave packet is normally incident on the interface, the wave vector associated with $\omega\!=\!\omega_{\mathrm{o}}$ is perpendicular to the interface [Fig.~\ref{Fig:NormalVsObliqueIncidence}(a)], whereas all other temporal frequencies $\omega$ are not. Two quantities are conserved across the interface: the transverse momentum $k_{x}$ and the temporal frequency $\omega$. If the refractive indices of the media are $n_{1}$ and $n_{2}$, and the group indices of the incident and transmitted wave packets are $\widetilde{n}_{1}$ and $\widetilde{n}_{2}$, respectively, then it can be shown that:
\begin{equation}\label{Eq:NormalIncidenceLaw}
n_{1}(n_{1}-\widetilde{n}_{1})=n_{2}(n_{2}-\widetilde{n}_{2}).
\end{equation}
This law of refraction for ST wave packets was derived in paper (I) \cite{Yessenov21RefractionI} and explored experimentally in paper (II) \cite{Yessenov21RefractionII}; see also Ref.~\cite{Bhaduri20NP}. The quantity $n(n-\widetilde{n})$ is related to the curvature of the spatio-temporal spectrum $\tfrac{\omega-\omega_{\mathrm{o}}}{\omega_{\mathrm{o}}}\!=\!\tfrac{k_{x}^{2}/k_{\mathrm{o}}^{2}}{2n(n-\widetilde{n})}$ when projected onto the $(k_{x},\tfrac{\omega}{c})$-plane. We have denoted this newly identified refractive invariant the `spectral curvature'. The law of refraction in Eq.~\ref{Eq:NormalIncidenceLaw} thus expresses the invariance of the spectral curvature across the planar interface at normal incidence.

At \textit{oblique} incidence, the wave vector associated with $\omega_{\mathrm{o}}$ is incident at an angle $\phi_{1}$ with respect to the normal to the interface [Fig.~\ref{Fig:NormalVsObliqueIncidence}(b)]. The group indices of the incident and transmitted ST wave packets satisfy a modified law \cite{Bhaduri20NP,Yessenov21RefractionI}:
\begin{equation}\label{Eq:ObliqueIncidenceLaw}
n_{1}(n_{1}-\widetilde{n}_{1})\cos^{2}{\phi_{1}}=n_{2}(n_{2}-\widetilde{n}_{2})\cos^{2}{\phi_{2}},
\end{equation}
where $\phi_{2}$ is obtained from Snell's law $n_{1}\sin{\phi_{1}}\!=\!n_{2}\sin{\phi_{2}}$. Equation~\ref{Eq:ObliqueIncidenceLaw} indicates the invariance of a new spectral curvature $n(n-\widetilde{n})\cos^{2}{\phi}$ at oblique incidence. The two curves corresponding to Eq.~\ref{Eq:NormalIncidenceLaw} and Eq.~\ref{Eq:ObliqueIncidenceLaw} intersect at the luminal point $\widetilde{n}_{1}\!=\!n_{1}$ and $\widetilde{n}_{2}\!=\!n_{2}$, where the ST wave packets revert to plane-wave pulses [Fig.~\ref{Fig:ObliqueIncidenceData}], and Eq.~\ref{Eq:ObliqueIncidenceLaw} reverts back to Eq.~\ref{Eq:NormalIncidenceLaw} at normal incidence $\phi_{1}\!=\!\phi_{2}\!=\!0$.

The rationale for the modified oblique-incidence spectral curvature can be elucidated by reference to Fig.~\ref{Fig:NormalVsObliqueIncidence}(b) and Fig.~\ref{Fig:Configuration}. The spectral support domain of the incident ST wave packet in the coordinate system $(x,z)$ -- which is aligned with the propagation direction of the wave packet --  lies at the intersection of the light-cone with the plane $\mathcal{P}_{\mathrm{B}}(\theta)$ [Fig.~\ref{Fig:Configuration}(a)]. At normal incidence, the projection of this spectrum onto the $(k_{x},\tfrac{\omega}{c})$-plane is invariant across the interface \cite{Bhaduri19Optica,Bhaduri20NP}. At oblique incidence, the transverse wave number $k_{x}$ is \textit{not} parallel to the interface, and therefore is \textit{not} conserved. However, in the coordinate system $(x',z')$ that is aligned with the interface rather than with the propagation direction of the ST wave packet, the transverse wave number $k_{x}'\!=\!k_{x}\cos{\phi}-k_{z}\sin{\phi}$ is conserved. In this coordinate system, the spectral support domain is rotated an angle $\phi_{1}$ around the $\tfrac{\omega}{c}$-axis [Fig.~\ref{Fig:Configuration}(b)], and the projection of this new spectrum onto the $(k_{x}',\tfrac{\omega}{c})$-plane is invariant across the interface [Fig.~\ref{Fig:Configuration}(c)]. It can be shown that $k_{x}\cos{\phi}$ is invariant to first order (in the small-angle approximation $\Delta k_{x}\!\ll\!k_{\mathrm{o}}$), and the new invariant spectral curvature at oblique incidence is $n(n-\widetilde{n})\cos^{2}{\phi}$. Because the opening angle of the light-cone changes from $\arctan{(n_{1})}$ in the first medium to $\arctan{(n_{2})}$ in the second, the invariance of the spectral projection onto the $(k_{x}',\tfrac{\omega}{c})$-plane produces a new spectral support domain for the transmitted wave packet on the surface of the light-cone [Fig.~\ref{Fig:Configuration}(d)]. A rotation through an angle $\phi_{2}$ (rather than $\phi_{1}$) around the $\tfrac{\omega}{c}$-axis returns the spectral support domain to the $(x,z)$ coordinate system aligned with the propagation direction of the transmitted wave packet [Fig.~\ref{Fig:Configuration}(e)]. 

\section{Confirming the law of refraction at oblique incidence}

\begin{figure}[t!]
\centering
\includegraphics[width=8.6cm]{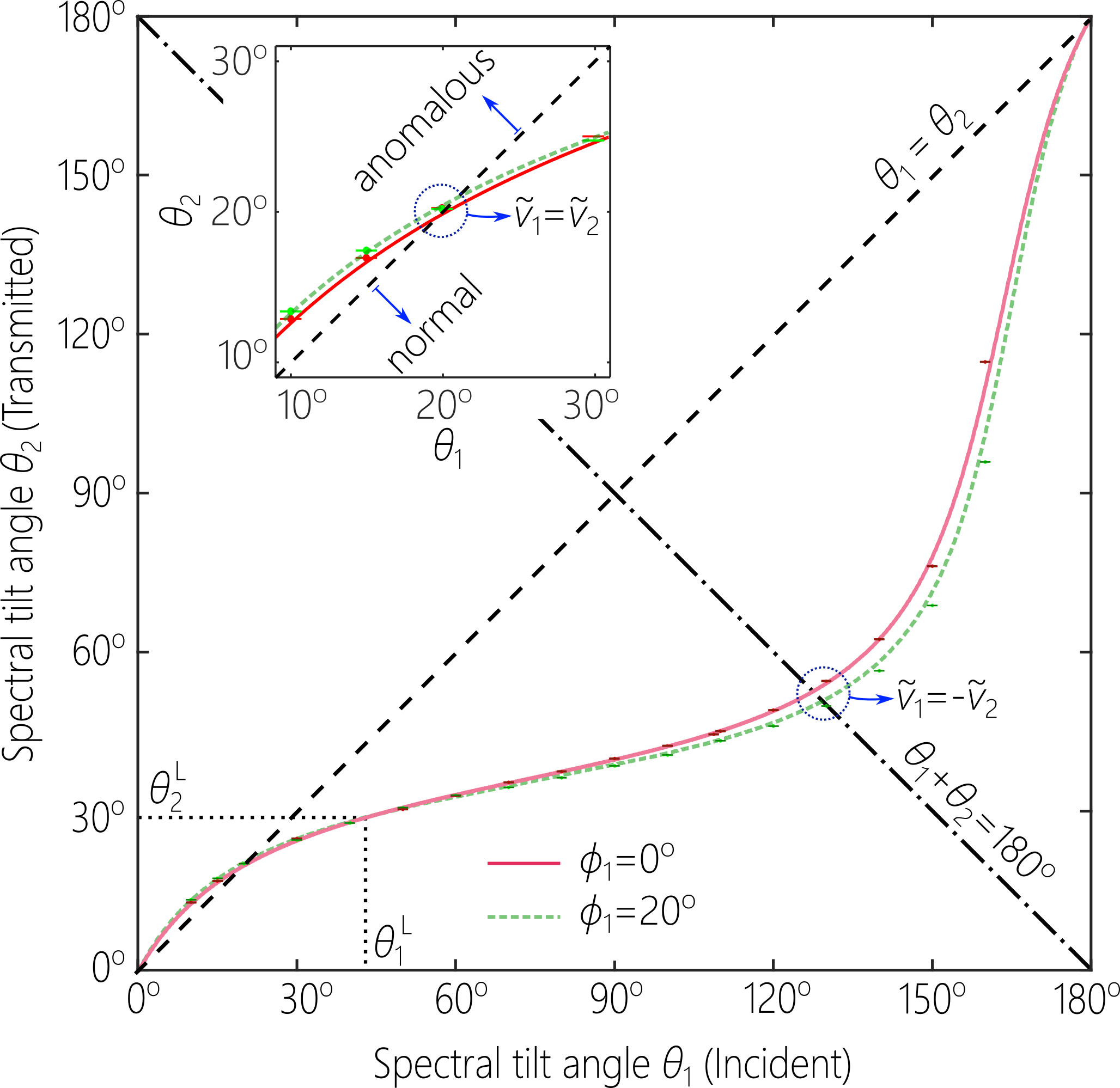}
\caption{Verification of the law of refraction for ST wave packets at oblique incidence (Eq.~\ref{Eq:ObliqueIncidenceLaw}) for $\phi_{1}\!=\!20^{\circ}$, compared to normal incidence (Eq.~\ref{Eq:NormalIncidenceLaw}) $\phi_{1}\!=\!0^{\circ}$. The inset highlights the transition from normal refraction to anomalous refraction. We also identify the positions along the curves corresponding to group-velocity invariance $\widetilde{v}\!=\!\widetilde{v}_{1}$ and group-velocity inversion $\widetilde{v}_{2}\!=\!-\widetilde{v}_{1}$.}
\label{Fig:ObliqueIncidenceData}
\end{figure}

To verify the law of refraction in Eq.~\ref{Eq:ObliqueIncidenceLaw} for baseband ST wave packets at oblique incidence, we make use of the interferometric procedure outlined in paper (II); see also \cite{Kondakci19NC,Bhaduri19Optica}. The planar slabs used in our measurements are placed in the common path of the co-aligned ST wave packet and the reference pulse, and we tilt the sample with respect to the incident wave packets. We plot the measurement results in Fig.~\ref{Fig:ObliqueIncidenceData}. Here we fix the angle of incidence $\phi_{1}$ and sweep the spectral tilt angle $\theta_{1}$ of the ST wave packet incident from free space onto a 5-mm-thick layer of sapphire. For each value of $\theta_{1}$, we measure the group delay incurred by the wave packet across the layer, from which we estimate $\theta_{2}$. The measurements are carried out at normal incidence $\phi_{1}\!=\!0$ and at oblique incidence $\phi_{1}\!=\!20^{\circ}$. The measurements verify for the first time the change in the law of refraction for ST wave packets with respect to that at normal incidence, which was demonstrated in \cite{Bhaduri20NP}. It is clear that the two curves for normal and oblique incidence in Fig.~\ref{Fig:ObliqueIncidenceData} intersect at the point corresponding to the luminal condition $(\theta_{1},\theta_{2})\!=\!(\theta_{1}^{\mathrm{L}},\theta_{2}^{\mathrm{L}})$. The curve for oblique incidence is shifted \textit{below} that for normal incidence in the superluminal regime ($\theta_{1}\!>\!\theta_{1}^{\mathrm{L}}$ and $\widetilde{v}\!>\!c/n$) and \textit{above} it in the subluminal regime ($\theta_{1}\!<\!\theta_{1}^{\mathrm{L}}$ and $\widetilde{v}\!<\!c/n$). We next explore the far-reaching ramifications of this change.

\section{Dependence of the group velocity on the angle of incidence}

\begin{figure}[t!]
\centering
\includegraphics[width=8.3cm]{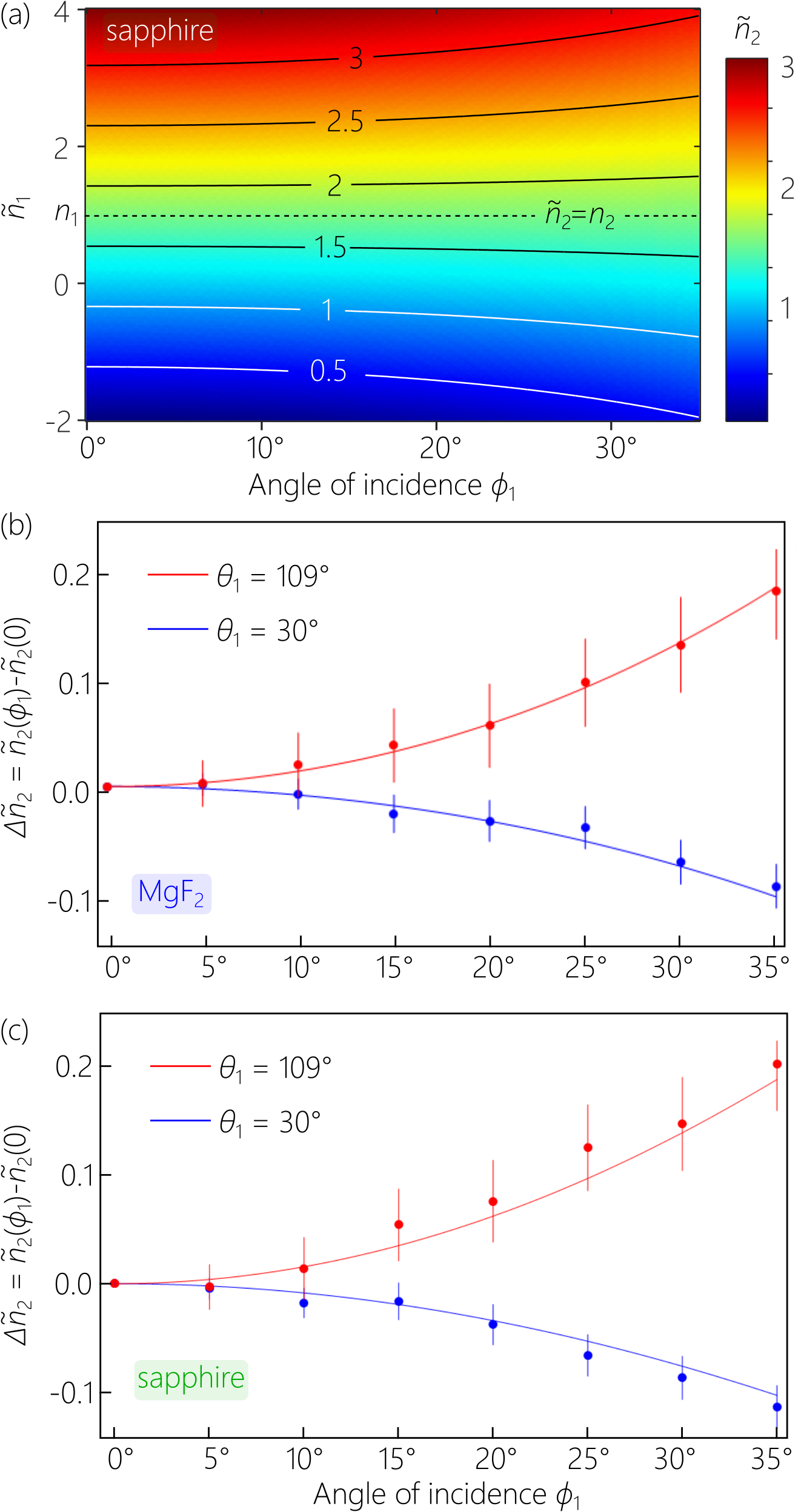}
\caption{(a) Calculated group index $\widetilde{n}_{2}$ of the transmitted ST wave packet as a function of the angle of incidence $\phi_{1}$ and the group index $\widetilde{n}_{1}$ of the incident ST wave packet from free space ($n_{1}\!=\!1$) onto sapphire ($n_{2}\!=\!1.76$). The dashed line is the locus of luminal plane-wave pulses whose spatial and temporal DoFs are separable; here $\widetilde{n}_{2}\!=\!n_{2}\!=\!1.76$ is independent of $\phi_{1}$. (b) Measured change in $\widetilde{n}_{2}$ at oblique incidence onto MgF$_2$ ($n_{2}\!=\!1.38$) from free space with respect to normal incidence, $\Delta\widetilde{n}_{2}\!=\!\widetilde{n}_{2}(\phi_{1})-\widetilde{n}_{2}(0)$. Measurements are carried out for a ST wave packet in the subluminal regime $\theta_{1}\!=\!30^{\circ}$ ($\widetilde{v}\!=\!0.58c$) and in the superluminal regime $\theta_{1}\!=\!109^{\circ}$ ($\widetilde{v}\!=\!-2.9c$). (c) Same as (a) for incidence from free space onto sapphire.}
\label{Fig:ChangeInGroupIndex}
\end{figure}

Figure~\ref{Fig:ObliqueIncidenceData} yields the relationship between the group indices $\widetilde{n}_{1}\!=\!\cot{\theta_{1}}$ and $\widetilde{n}_{2}\!=\!\cot{\theta_{2}}$ for the incident and refracted ST wave packets, respectively, at a \textit{fixed} incident angle $\phi_{1}$. We next examine the change in the group index $\widetilde{n}_{2}$ for the transmitted wave packet as we vary the incidence angle $\phi_{1}$ while holding fixed the group index $\widetilde{n}_{1}$ for the incident wave packet. In Fig.~\ref{Fig:ChangeInGroupIndex}(a), we plot the calculated $\widetilde{n}_{2}$ while varying both $\widetilde{n}_{1}$ and $\phi_{1}$ for incidence from free space $n_{1}\!=\!1$ onto sapphire $n_{2}\!=\!1.76$. This calculation highlights the above-mentioned `memory' effect. The group index $\widetilde{n}_{2}$ of the transmitted wave packet depends not only the refractive index of the second medium $n_{2}$, but also on the characteristics of the incident wave packet: its group index $\widetilde{n}_{1}$ and incident angle $\phi_{1}$. The luminal condition for the incident ST wave packet is $\widetilde{n}_{1}\!=\!n_{1}\!=\!1$, in which case this wave packet is simply a plane-wave pulse, and the refracted wave packet is also a luminal plane-wave pulse $\widetilde{n}_{2}\!=\!n_{2}\!=\!1.76$. For subluminal incidence ST wave packets $\widetilde{n}_{1}\!>\!n_{1}$, $\widetilde{n}_{2}$ \textit{decreases} with $\phi_{2}$; i.e., the group velocity of the transmitted subluminal wave packet increases with $\phi_{1}$ in the higher-index medium. Conversely, for superluminal incident ST wave packets $\widetilde{n}_{1}\!<\!n_{1}$, $\widetilde{n}_{2}$ \textit{increases} with $\phi_{1}$; i.e., the group velocity of the transmitted superluminal wave packet increases with $\phi_{1}$ in the higher-index medium. In the case of a conventional wave packet whose spatial and temporal DoFs are separable, $\widetilde{n}_{2}$ is independent of $\phi_{1}$.

This behavior can be understood on the basis of Eq.~\ref{Eq:ObliqueIncidenceLaw}. Substituting for $\phi_{2}$ in terms of $\phi_{1}$ from Snell's law yields:
\begin{equation}\label{Eq:IndexInSecondMaterialWIthPhi1}
\frac{\widetilde{n}_{2}(\phi_{1})}{n_{2}}=1+\left(\frac{\widetilde{n}_{1}}{n_{1}}-1\right)\eta(\phi_{1}),
\end{equation}
where the factor $\eta(\phi_{1})$ depends on the indices $n_{1}$ and $n_{2}$ through
\begin{equation}
\eta(\phi_{1})=\frac{\cos^{2}{\phi_{1}}}{(\frac{n_{2}}{n_{1}})^{2}-\sin^{2}{\phi_{1}}}.
\end{equation}
The behavior of $\eta(\phi_{1})$ depends on the ratio $n_{1}/n_{2}$. When $n_{1}\!<\!n_{2}$, $\eta(\phi_{1})$ drops monotonically from its initial value to $0$ at $\phi_{1}\!=\!90^{\circ}$. Because the term $(\tfrac{\widetilde{n}_{1}}{n_{1}}-1)$ is positive in the subluminal regime $\widetilde{n}_{1}\!>\!n_{1}$, the decrease of $\eta(\phi_{1})$ with $\phi_{1}$ results in a \textit{decrease} of $\widetilde{n}_{2}(\phi_{1})$ with $\phi_{1}$. In the superluminal regimes, the term $(\tfrac{\widetilde{n}_{1}}{n_{1}}-1)$ is negative, and $\widetilde{n}_{2}(\phi_{1})$ \textit{increase} with $\phi_{1}$. On the other hand, when $n_{1}\!>\!n_{2}$, $\eta(\phi_{1})$ increases monotonically with $\phi_{1}$ and reaches a singularity when $\sin{\phi_{1}}\!=\!n_{2}/n_{1}$, corresponding to the critical angle (total internal reflection) at the interface. Consequently, the opposite trends for $\widetilde{n}_{2}$ with $\phi_{1}$ ensue.

Our measurements confirm this predicted behavior. In Fig.~\ref{Fig:ChangeInGroupIndex}(b,c) we plot the change in the refracted group index $\widetilde{n}_{2}$ with $\phi_{1}$ with respect to that at normal incidence, $\Delta\widetilde{n}_{2}\!=\!\widetilde{n}_{2}(\phi_{1})-\widetilde{n}_{2}(0)$, for incidence from free space onto two materials: MgF$_2$ in Fig.~\ref{Fig:ChangeInGroupIndex}(b) and sapphire in Fig.~\ref{Fig:ChangeInGroupIndex}(c). For each material, we carry out measurements with ST wave packets synthesized in free space in the subluminal ($\theta_{1}\!<\!45^{\circ}$) and superluminal ($\theta_{1}\!>\!45^{\circ}$) regimes. We note several general observations about the results that are expected from our analysis. First, $\widetilde{n}_{2}(\phi_{1})$ deviates monotonically from the normal-incidence value $\widetilde{n}_{2}(0)$. Second, the group index drops at oblique incidence $\widetilde{n}_{2}(\phi_{1})\!<\!\widetilde{n}_{2}(0)$ in the subluminal regime, and increases $\widetilde{n}_{2}(\phi_{1})\!>\!\widetilde{n}_{2}(0)$ in the superluminal regime. That is, the group velocity of a refracted \textit{subluminal} ST wave packet \textit{increases} with incident angle in the higher-index medium. Conversely, the group velocity of a refracted \textit{superluminal} ST wave packet \textit{decreases} with $\phi_{1}$. These features are critical for the blind synchronization scheme we investigate below.

\section{Impact of the angle of incidence on group-velocity invariance and inversion}

The law of refraction at normal incidence (Eq.~\ref{Eq:NormalIncidenceLaw}) predicts three phenomena that occur for any pair of media: (1) group velocity invariance $\widetilde{v}_{2}\!=\!\widetilde{v}_{1}$, which occurs when $\widetilde{n}_{1}\!=\!\widetilde{n}_{\mathrm{th}}\!=\!n_{1}+n_{2}$; (2) anomalous refraction whereby $\widetilde{v}_{2}\!>\!\widetilde{v}_{1}$ when $n_{2}\!>\!n_{1}$, which occurs when $\widetilde{n}_{1}\!>\!\widetilde{n}_{\mathrm{th}}$; and (3) group-velocity inversion $\widetilde{v}_{2}\!=\!-\widetilde{v}_{1}$, which occurs when $\widetilde{n}_{1}\!=\!n_{1}-n_{2}$; see Fig.~\ref{Fig:ObliqueIncidenceData}. These three phenomena were confirmed experimentally in Ref.~\cite{Bhaduri20NP} and in more detail in paper (II) \cite{Yessenov21RefractionII}. Despite the difference between the law of refraction at oblique incidence (Eq.~\ref{Eq:ObliqueIncidenceLaw}) from that at normal incidence (Eq.~\ref{Eq:NormalIncidenceLaw}), these three phenomena are still realizable, albeit with modifications to the conditions to be satisfied.

\begin{figure}[t!]
\centering
\includegraphics[width=8.6cm]{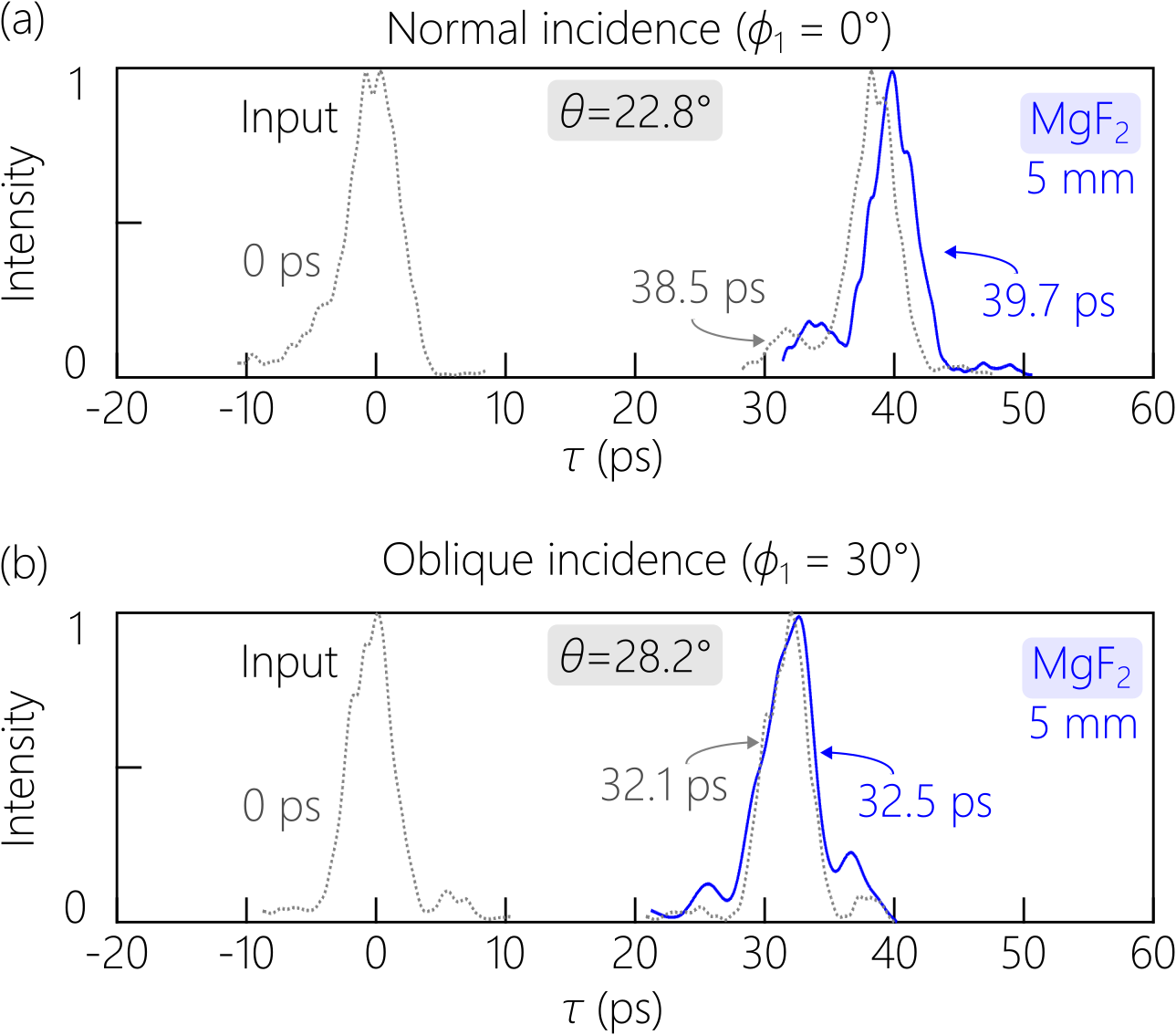}
\caption{Impact of oblique incidence on the condition for realizing group-velocity invariance. (a) Normal incidence, $\phi_{1}\!=\!0$. We plot the on-axis $x\!=\!0$ input-pulse profile $I(z\!=\!0;\tau)$ on the left, and the output pulse profiles $I(L;\tau)$ after traversing $L\!=\!5$~mm of free space (dotted curve) and MgF$_2$ (solid curve) at normal incidence for a ST wave packet with $\theta_{1}\!=\!22.8^{\circ}$ ($\widetilde{v}\!=\!0.42c$) synthesized in free space. (b) Same as (a) for oblique incidence at $\phi_{1}\!=\!30^{\circ}$. Here the group delays traversing equal lengths of free space and MgF$_2$ are equal but have a smaller overall delay than in (a), despite the longer distance traveled.}
\label{Fig:GroupVelocityInvariance}
\end{figure}

First, group-velocity \textit{invariance} occurs at oblique incidence when \cite{Yessenov21RefractionI}:
\begin{equation}
\widetilde{n}_{1}=\widetilde{n}_{\mathrm{th}}(\phi_{1})=\frac{n_{1}+n_{2}}{1+\frac{n_{1}}{n_{2}}\sin^{2}{\phi_{1}}},
\end{equation}
where $\widetilde{n}_{\mathrm{th}}(\phi_{1})\!<\!\widetilde{n}_{\mathrm{th}}(0)\!=\!n_{1}+n_{2}$. We plot in Fig.~\ref{Fig:GroupVelocityInvariance} measurements of the group delay incurred upon traversing equal lengths ($L\!=\!5$~mm) of free space ($n_{1}\!=\!1$) and MgF$_2$ ($n_{2}\!=\!1.38$). At normal incidence $\widetilde{n}_{\mathrm{th}}(0)\!=\!2.38$  corresponding to $\theta_{1}\!\approx\!22.8^{\circ}$ [Fig.~\ref{Fig:GroupVelocityInvariance}(a)], and at oblique incidence $\phi_{1}\!=\!30^{\circ}$ we have $\widetilde{n}_{\mathrm{th}}(30^{\circ})\!=\!1.86$, corresponding to $\theta\!\approx\!28.2^{\circ}$ [Fig.~\ref{Fig:GroupVelocityInvariance}(b)]. In both cases the ST wave packets accrue approximately the same delay in equal lengths of free space and MgF$_2$. Whereas anomalous refraction occurs at normal incidence when $\theta_{1}\!<\!22.8^{\circ}$, this regime is expanded for oblique incidence to $\theta_{1}\!<\!28.2^{\circ}$ at $\phi_{1}\!=\!30^{\circ}$.

\begin{figure}[t!]
\centering
\includegraphics[width=8.6cm]{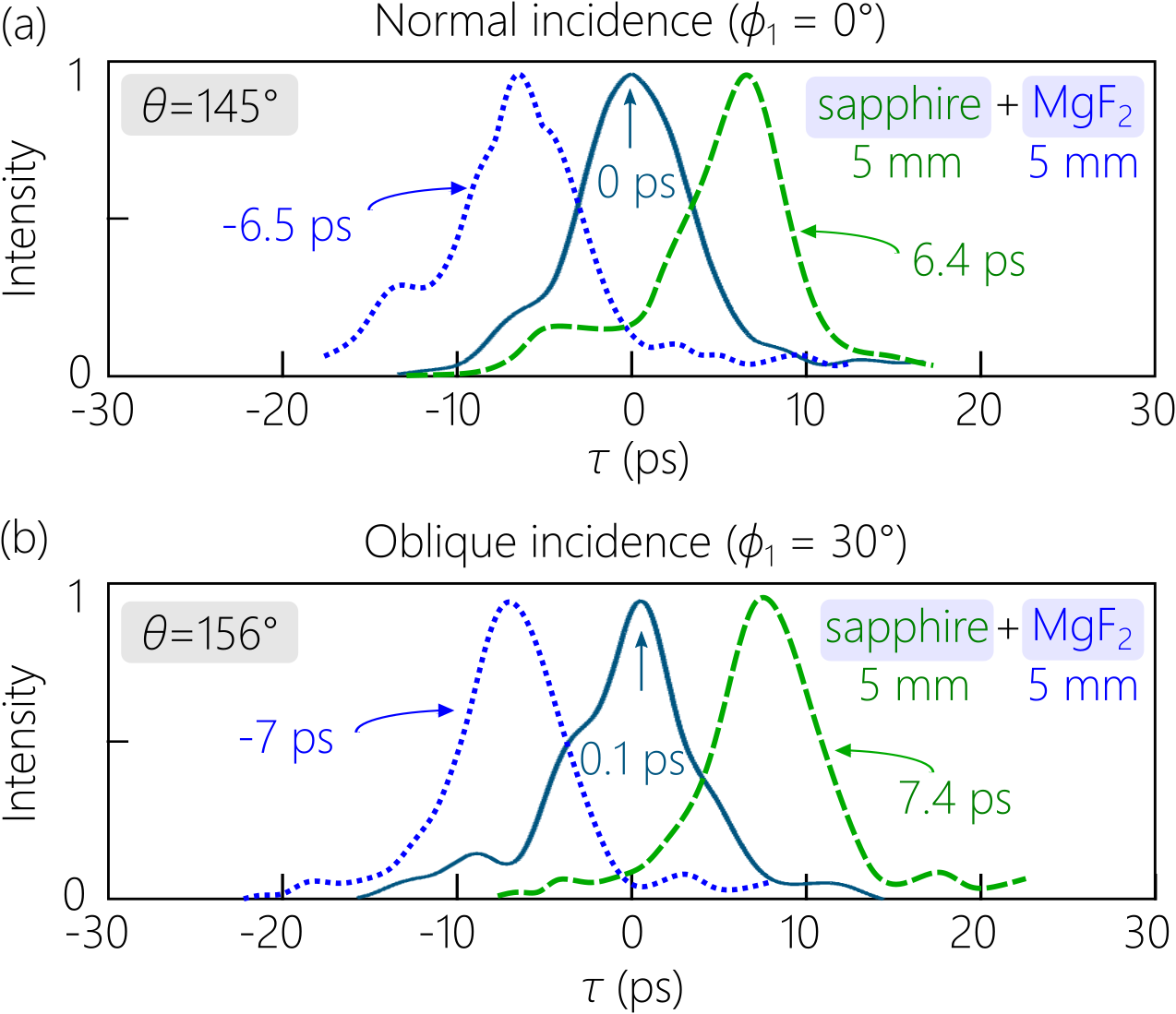}
\caption{Impact of oblique incidence on group-velocity inversion. (a) At normal incidence ($\phi_{1}\!=\!0$), a ST wave packet with $\theta_{0}\!=\!145^{\circ}$ traverses a 5-mm-thick layer of MgF$_2$ with a negative group delay (dashed curve; $\widetilde{v}_{1}\!=\!-2.6c$ and $\theta_{1}\!=\!111^{\circ}$ in MgF$_2$) and a 5-mm-thick layer of sapphire with a positive group delay (dotted curve; $\widetilde{v}_{2}\!=\!2.6c$ and $\theta_{1}\!=\!69^{\circ}$ in sapphire), corresponding to group delay inversion. When the ST wave packet is incident from free space on a bilayer of MgF$_2$ and sapphire, the total group delay is zero (solid curve), corresponding to group-delay cancellation. (b) Same as (a) for oblique incidence at $\phi_{1}\!=\!30^{\circ}$. The incident ST wave packet from free space has $\theta_{0}\!=\!156^{\circ}$. The delays incurred by the two pulses in the individual layers in larger than at normal incidence because the distance traveled is larger at oblique incidence and the wave packet is in the normal refraction regime.}
\label{Fig:GroupVelocityInversion}
\end{figure}

Second, group-velocity \textit{inversion} occurs at oblique incidence when \cite{Yessenov21RefractionI}:
\begin{equation}
\widetilde{n}_{1}(\phi_{1})=\frac{n_{1}-n_{2}}{1-\frac{n_{1}}{n_{2}}\sin^{2}{\phi_{1}}},
\end{equation}
where $\widetilde{n}_{1}(\phi_{1})\!<\!\widetilde{n}_{1}(0)\!=\!n_{1}-n_{2}$ when $n_{2}\!>\!n_{1}$. We plot in Fig.~\ref{Fig:GroupVelocityInversion} the measured group delay incurred upon traversing equal lengths ($L\!=\!5$~mm) of sapphire and MgF$_2$ separately, and after traversing a bilayer of the two media. The required group index for a ST wave packet in free space that experiences group-velocity invariance in sapphire and MgF$_2$ is $\widetilde{n}_{0}\!\approx\!-1.43$ ($\theta_{0}\!\approx\!145^{\circ}$); see Fig.~\ref{Fig:GroupVelocityInversion}(a). At oblique incidence ($\phi_{1}\!=\!30^{\circ}$), the required free-space group index to realize group-velocity inversion is $\widetilde{n}_{0}\approx\!-2.25$ ($\theta_{0}\!\approx\!156^{\circ}$); see Fig.~\ref{Fig:GroupVelocityInversion}(b). In both cases, the group delays measured in the individual layers are approximately equal in magnitude but opposite in sign. Consequently, the group delay in the bilayer almost vanishes as shown.

\section{Blind synchronization}

\subsection{Concept of blind synchronization using ST wave packets}

\begin{figure}[t!]
\centering
\includegraphics[width=7cm]{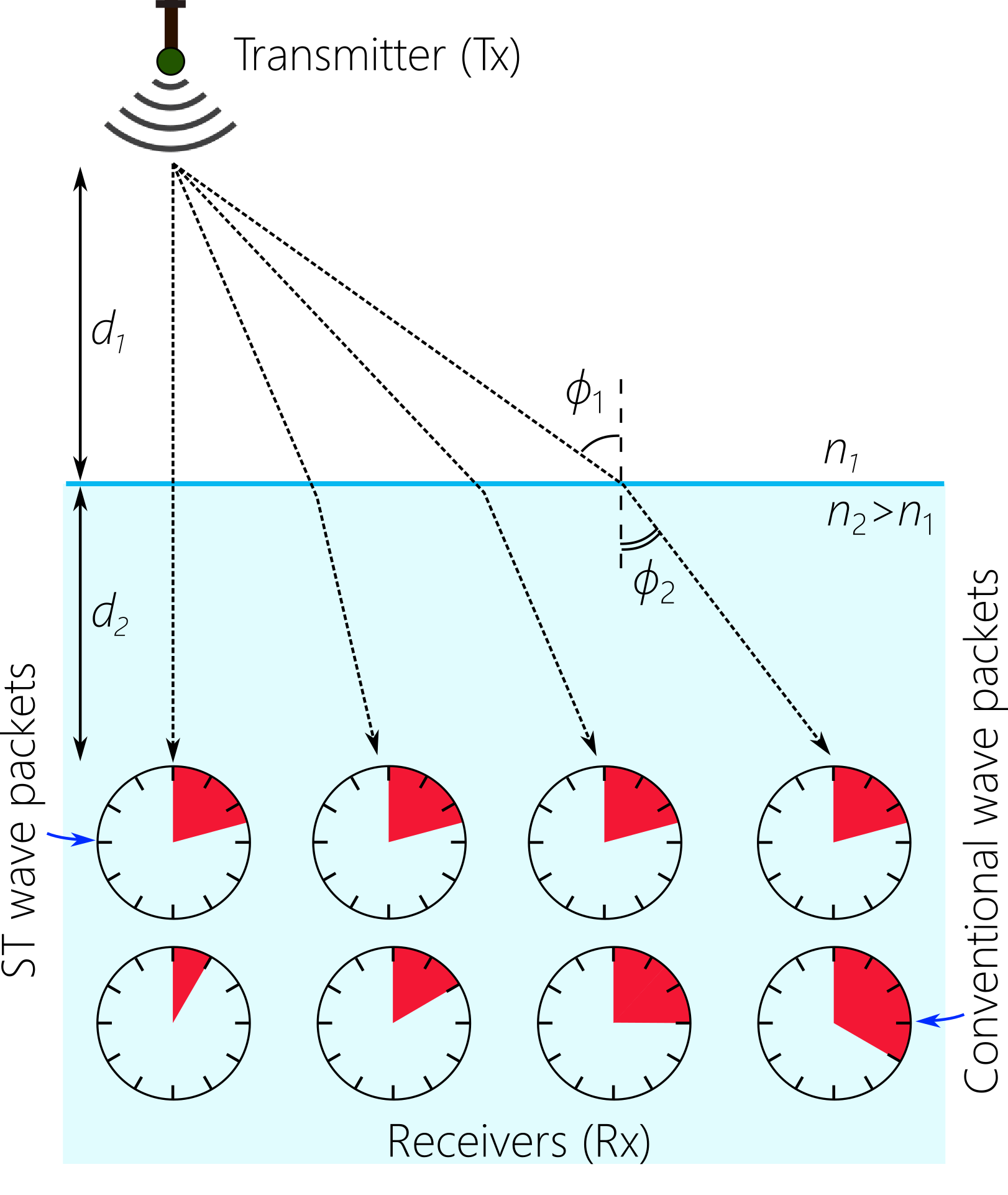}
\caption{Concept of blind synchronization. We depict the configuration for the blind synchronization of multiple receivers (Rx) receiving pulses from a transmitter (Tx). All the receivers are at the same depth $d_{2}$ beneath the interface, but their locations are otherwise not known \textit{a priori}. When using conventional wave packets, the group delay increases as the physical path length from Tx to Rx increases, which occurs when the incident angle $\phi_{1}$ at the interface increases. By employing ST wave packets instead, the group delay accrued along the different pathways from Tx to Rx can be held constant because the group velocity of the transmitted wave packet \textit{increases} with $\phi_{1}$, thereby compensating for the longer path length covered.}
\label{Fig:SynchronizationConfiguration}
\end{figure}

Consider the configuration shown in Fig.~\ref{Fig:SynchronizationConfiguration}. A transmitter (Tx) in the first medium at a height $d_{1}$ above the interface directs an ST wave packet at different angles to receivers (Rx) at different locations but at the same depth $d_{2}$ beneath the interface in the second medium. The optical delay in the first medium from Tx to the interface is $\tau_{1}(\phi_{1})\!=\!\tau_{1}(0)/\cos{\phi_{1}}$, where $\tau_{1}(0)\!=\!d_{1}\widetilde{n}_{1}/c$ is the group delay at normal incidence. The group velocity of the ST wave packet is the same in all directions independently of $\phi_{1}$ because the medium is homogeneous and isotropic. Therefore, the group delay $\tau_{1}(\phi_{1})$ increases with $\phi_{1}$. The group delay in the second medium is $\tau_{2}(\phi_{1})\!=\!d_{2}\widetilde{n}_{2}(\phi_{1})/(c\cos{\phi_{2}})$, where $\widetilde{n}_{2}(\phi_{1})$ is the group index of the transmitted wave packet, determined from Eq.~\ref{Eq:ObliqueIncidenceLaw}. In contrast to the incident wave packet, the group velocity of the transmitted wave packet depends on $\phi_{1}$. 

The relative group delay between obliquely and normally incident ST wave packets in the first medium is:
\begin{equation}
\Delta\tau_{1}=\tau_{1}(\phi_{1})-\tau_{1}(0)=\frac{d_{1}}{c}\widetilde{n}_{1}\left(\frac{1}{\cos{\phi_{1}}}-1\right),
\end{equation}
and in the second medium is:
\begin{equation}
\Delta\tau_{2}=\tau_{2}(\phi_{1})-\tau_{2}(0)=\frac{d_{2}}{c}\left(\frac{\widetilde{n}_{2}(\phi_{1})}{\cos{\phi_{2}}}-\widetilde{n}_{2}(0)\right).
\end{equation}
We define the total relative group delay $\Delta\tau\!=\!\Delta\tau_{1}+\Delta\tau_{2}\!=\!\tau(\phi_{1})-\tau(0)$, where $\tau(\phi_{1})\!=\!\tau_{1}(\phi_{1})+\tau_{2}(\phi_{1})$ is the total group delay incurred from Tx to Rx across the interface. Blind synchronization requires $\Delta\tau\!=\!0$ independently of $\phi_{1}$, such that the ST wave packet arrives simultaneously at all receivers at the depth $d_{2}$ despite the different physical distances covered along the different trajectories.

We first ascertain whether realizing $\Delta\tau\!=\!0$ is physically feasible. Note that $\Delta\tau_{1}$ in the first medium is always positive, because the group delay along the longer oblique path is always larger than that along the shorter normal path. Blind synchronization therefore requires that $\Delta\tau_{2}$ in the second medium be \textit{negative}. That is, the group velocity of the transmitted wave packet must increase at oblique incidence to compensate for the longer propagation distances in both media. In other words, a necessary (but not sufficient) condition is that $\widetilde{v}_{2}(\phi_{1})\!>\!\widetilde{v}_{2}(0)$. From the analysis and measurements reported above, this scenario occurs for subluminal ST wave packets when $n_{2}\!>\!n_{1}$. We next consider whether the blind synchronization condition can be met quantitatively.

\begin{figure}[b!]
\centering
\includegraphics[width=8.6cm]{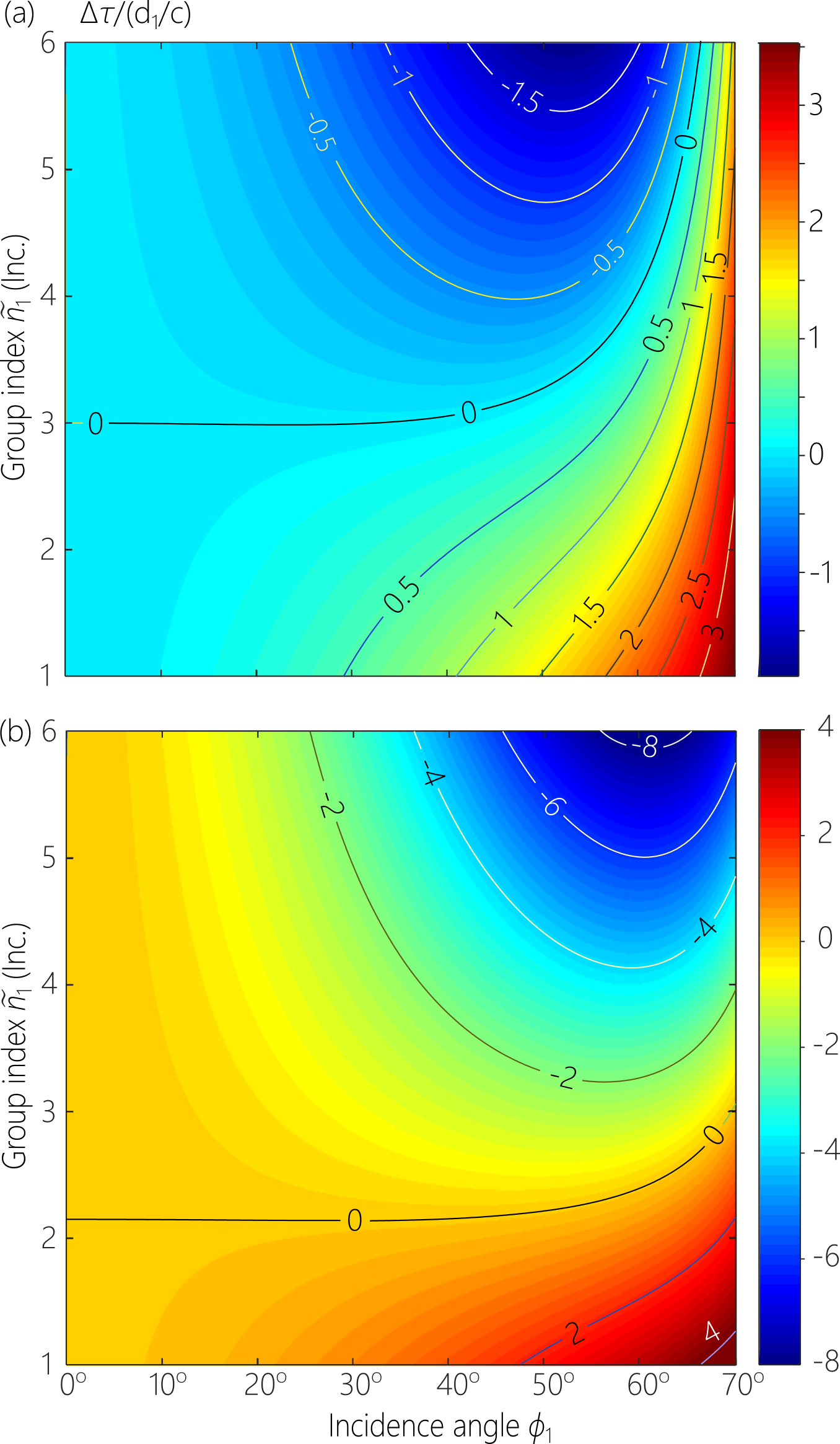}
\caption{Plots of the relative group delay $\Delta\tau$ (normalized with respect to $d_{1}/c$) as a function of the incident angle $\phi_{1}$ and the group index $\widetilde{n}_{1}$ of the incident ST wave packet: (a) $n_{1}\!=\!1$, $n_{2}\!=\!1.76$, and $d_{2}/d_{1}\!=\!5$; and (b) $n_{1}\!=\!1$, $n_{2}\!=\!2$, and $d_{2}/d_{1}\!=\!10$.}
\label{Fig:SynchronizationCalculations}
\end{figure}

\subsection{Calculated group delay of ST wave packets between transmitter and receivers}

The total relative group delay $\Delta\tau\!=\!\Delta\tau_{1}+\Delta\tau_{2}$ is given by:
\begin{eqnarray}\label{Eq:STWavePacketDelay}
\frac{\Delta\tau(\phi_{1})}{d_{1}/c}\!\!\!\!&=&\!\!\!\!n_{2}\frac{d_{2}}{d_{1}}\left\{\!\left(\!\frac{1}{\cos{\phi_{1}}}-1\right)+\left(\!\frac{\widetilde{n}_{1}}{n_{1}}\!-\!1\!\right)\left(\!\frac{\eta(\phi_{1})}{\cos{\phi_{2}}}-\eta(0)\!\right)\!\right\}\nonumber\\\!\!\!\!&+&\widetilde{n}_{1}\left(\frac{1}{\cos{\phi_{1}}}\!-\!1\!\right).
\end{eqnarray}
where we made use of the law of refraction in Eq.~\ref{Eq:ObliqueIncidenceLaw}, and we normalized $\Delta\tau$ with respect to $d_{1}/c$. We plot in Fig.~\ref{Fig:SynchronizationCalculations} calculations for $\Delta\tau$ as a function of the incident angle $\phi_{1}$ and the group index of the incident wave packet $\widetilde{n}_{1}$ in two different configurations. In the first configuration we have $n_{1}\!=\!1$, $n_{2}\!=\!1.76$, and $d_{2}/d_{1}\!=\!5$; and in the second configuration we have $n_{1}\!=\!1$, $n_{2}\!=\!2$, and $d_{2}/d_{1}\!=\!10$. In both cases we find the contour $\Delta\tau\!=\!0$ occurs at a specific group index $\widetilde{n}_{1}$ over a range of incident angles. Such ST wave packets arrive simultaneously at receivers located at the same depth within this angular range despite following paths of very different lengths, thereby realizing blind synchronization. We find in general that meeting the requirement for blind synchronization is more favorable when $d_{2}/d_{1}$ and/or $n_{2}/n_{1}$ increase.

\begin{figure}[t!]
\centering
\includegraphics[width=8.6cm]{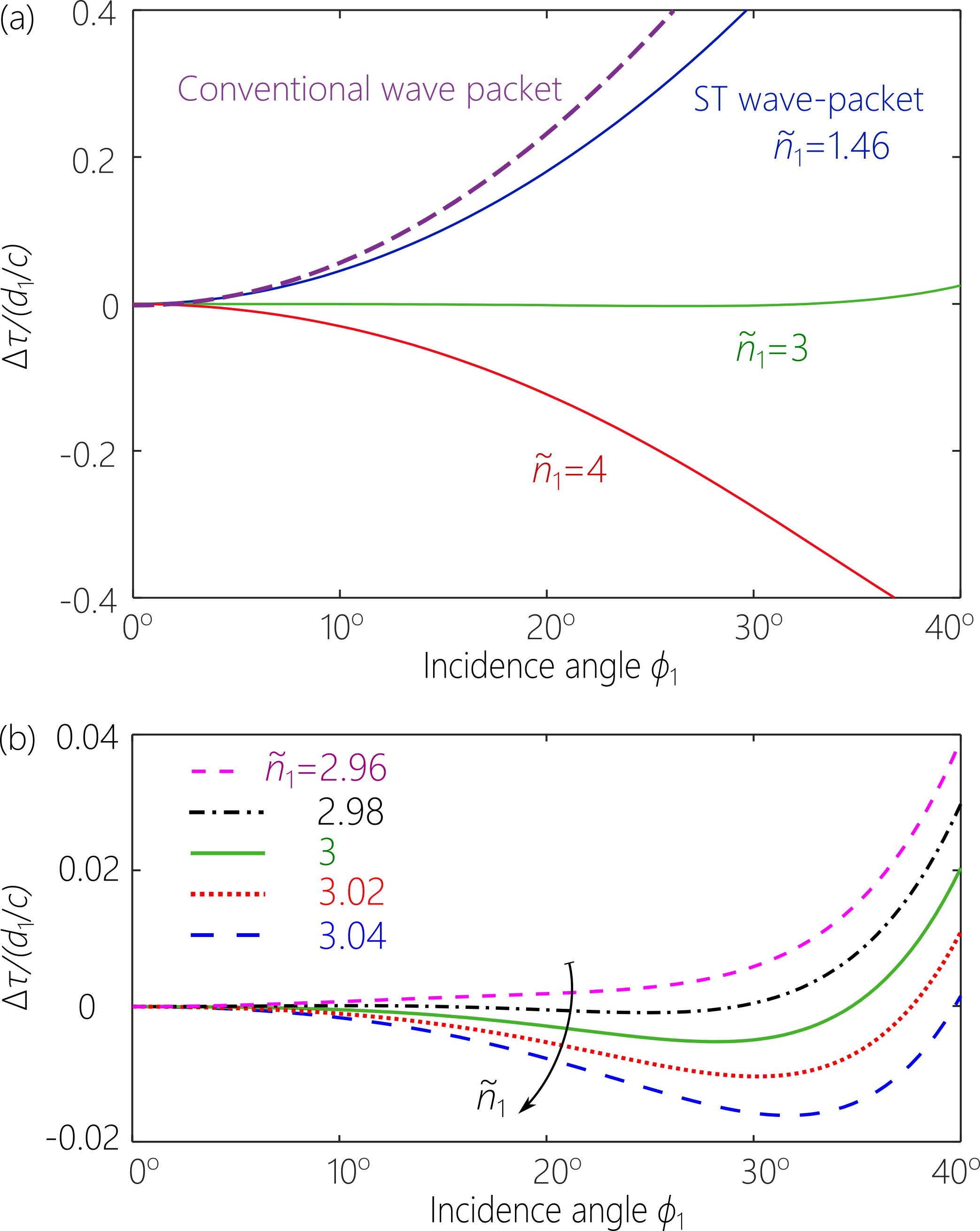}
\caption{(a) Comparison of the relative group delay $\Delta\tau$ (normalized with respect to $d_{1}/c$) of ST wave packets (solid curves; Eq.~\ref{Eq:STWavePacketDelay}) from Fig.~\ref{Fig:SynchronizationCalculations}(a) at $\widetilde{n}_{1}\!=\!1.46$, 3 and 4 with $\Delta\tau$ of a traditional pulse (dashed curve; Eq.~\ref{Eq:TraditionalPulseDelay}). (b) Same as (a), but with a reduced range for $\Delta\tau$ along the vertical axis.}
\label{Fig:SynchronizationComparison}
\end{figure}

For the sake of comparison, we plot in Fig.~\ref{Fig:SynchronizationComparison}(a) the relative group delay $\Delta\tau$ for ST wave packets having different group indices alongside that for a \textit{conventional} wave packet whose spatial and temporal DoFs are separable. The conventional wave packet propagates in the first medium at a group velocity $c/n_{1}$ and in the second at $c/n_{2}$ (assuming non-dispersive media). The relative group delay is given by:
\begin{equation}\label{Eq:TraditionalPulseDelay}
\frac{\Delta\tau_{\mathrm{conv}}(\phi_{1})}{d_{1}/c}=n_{1}\left(\frac{1}{\cos{\phi_{1}}}-1\right)+n_{2}\frac{d_{2}}{d_{1}}\left(\frac{1}{\cos{\phi_{2}}}-1\right).
\end{equation}
The relative group delay $\Delta\tau$ is always positive for conventional pulses because the pulses traveling longer distances inevitably accrues larger group delays. In contrast, varying $\widetilde{n}_{1}$ for ST wave packets changes $\Delta\tau$ from positive values (similarly to a conventional pulse where longer paths incur longer delays) to anomalously negative values (longer paths incur \textit{shorter} delays) passing through the desired target of blind synchronization $\Delta\tau\!=\!0$. In Fig.~\ref{Fig:SynchronizationComparison}(b) we highlight this transition from $\Delta\tau\!>\!0$ to $\Delta\tau\!<\!0$ as $\widetilde{n}_{1}$ is changed in the vicinity of $\widetilde{n}_{1}\!=\!3$.

\begin{figure}[t!]
\centering
\includegraphics[width=8.6cm]{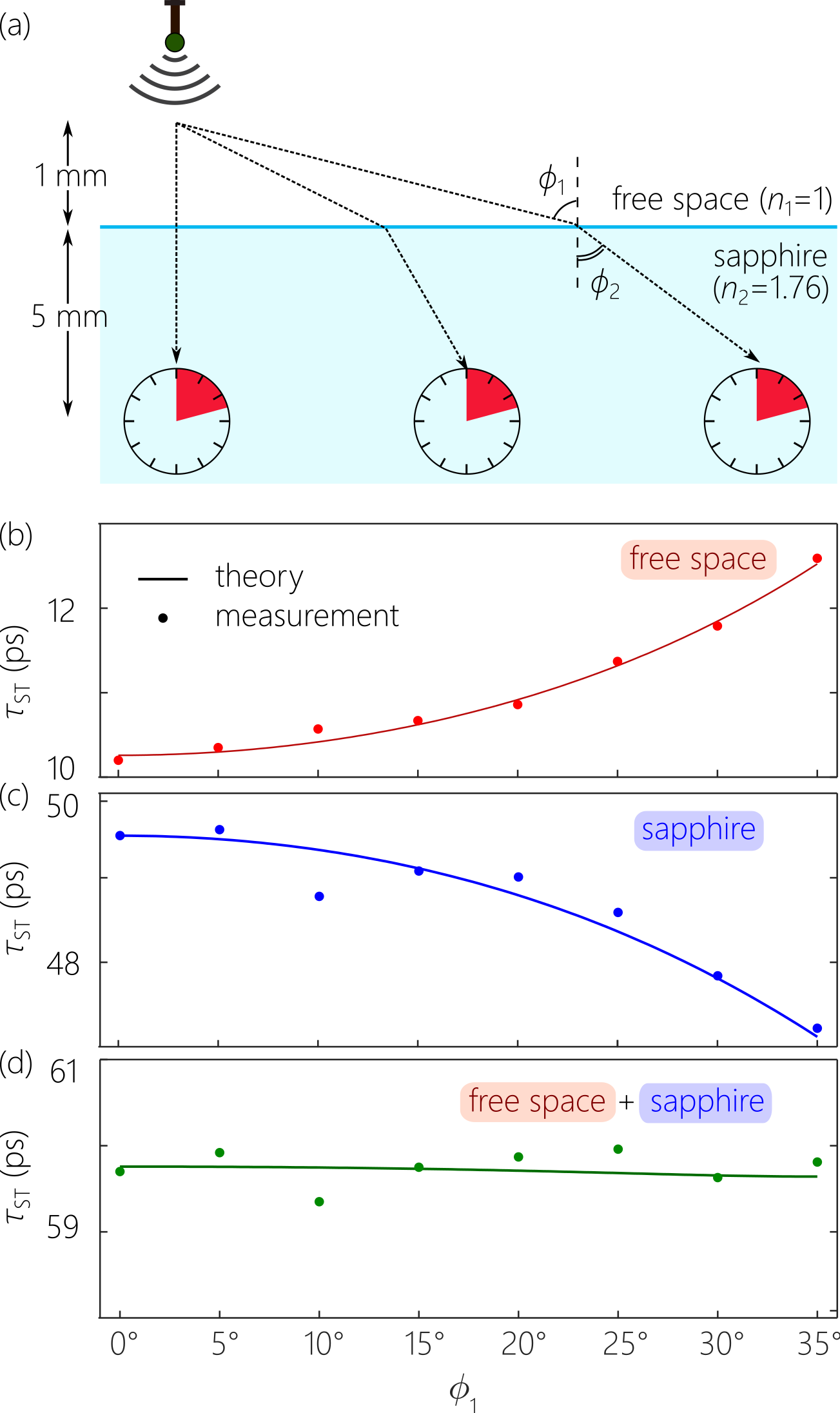}
\caption{Proof-of-principle realization of blind synchronization. (a) Schematic of the experimental configuration. (b) Measured group delay accrued over 1~mm of free space by a ST wave packet with $\widetilde{n}_{1}\!=\!3$. The delay increases with $\phi_{1}$. (c) Measured group delay over 5~mm of sapphire for the ST wave packet from (b). Here, the group delay \textit{decreases} with $\phi_{1}$. (d) Sum of the group delays from (b) and (c), showing an approximately constant value despite the different path lengths associated with different incident angles $\phi_{1}$.}
\label{Fig:BlindSynchronization}
\end{figure}

\subsection{Blind synchronization}

We plot in Fig.~\ref{Fig:BlindSynchronization} results for a proof-of-principle test of the blind synchronization effect. The overall system configuration is illustrated in Fig.~\ref{Fig:BlindSynchronization}(a). We consider the configuration in Fig.~\ref{Fig:SynchronizationCalculations}(a) for propagation from free space ($n_{1}\!=\!1$) to sapphire ($n_{2}\!=\!1.76$) with the propagation distances at normal incidence having the ratio $d_{2}/d_{1}\!=\!5$; we use here $d_{1}\!=\!1$~mm and $d_{2}\!=\!5$~mm. We measure the group delay of the ST wave packet in free space [Fig.~\ref{Fig:BlindSynchronization}(b)] and in the sapphire layer [Fig.~\ref{Fig:BlindSynchronization}(c)] separately. The separate delays for the oblique-incidence paths are measured by tilting the sapphire layer and in free space by extending the distance $d_{1}\!\rightarrow\!d_{1}/\cos{\phi_{1}}$ over which the delay is measured.

We synthesize a subluminal ST wave packet in free space with $\widetilde{n}_{1}\!=\!3$ ($\theta_{1}\!=\!18^{\circ}$). According to Fig.~\ref{Fig:SynchronizationCalculations}(a), blind synchronization $\Delta\tau\!=\!0$ is realized by such a wave packet over a range of incidence angles $\phi_{1}\!=\!0\rightarrow\pm35^{\circ}$. We plot in Fig.~\ref{Fig:BlindSynchronization}(b) the measured group delay in free space over distances extending from $d_{1}\!=\!1$~mm (corresponding to $\phi_{1}\!=\!0^{\circ}$) to $d_{1}\!=\!1.2$~mm (corresponding to $\phi_{1}\!=\!35^{\circ}$), with $d_{1}(\phi_{1})\!=\!d_{1}/\cos{\phi_{1}}$. We next plot in Fig.~\ref{Fig:BlindSynchronization}(c) the measured group delay in the sapphire layer as we increase the angle of incidence. Here, the group delay \textit{decreases} with the increases distance traveled in the layer at larger $\phi_{1}$. As shown in Fig.~\ref{Fig:ChangeInGroupIndex}, this is a general feature at oblique incidence for subluminal ST wave packets when $n_{2}\!>\!n_{1}$. For this particular ST wave packet with $\widetilde{n}_{1}\!=\!3$, the decrease in delay in the layer (increase in group velocity) with $\phi_{1}$ is sufficient to counterbalance the increase in delay in free space \textit{and} the increase in propagation distance in the sapphire layer, such that the total delays in free space and the layer are approximately equal [Fig.~\ref{Fig:BlindSynchronization}(d)] for all trajectories corresponding to incident angles from $\phi_{1}\!=\!0^{\circ}$ to $\phi_{1}\!=\!35^{\circ}$.

\section{Discussion and Conclusions}

The proof-of-principle experimental demonstration of blind synchronization described here extends our recent report on \textit{isochronous} ST wave packets \cite{Motz21arxiv}. These are wave packets that traverse a material layer after accruing a fixed group delay independently of the angle of incidence, despite traveling larger distances in the layer ($L/\cos{\phi_{2}}$, where $L$ is the layer thickness). This can be seen as a special case of blind synchronization where we set $d_{1}\!=\!0$. An increase in group velocity with incident angle for isochronous ST wave packets is required to compensate for the increase in propagation distance within the layer. This is achieved, as we have done here for blind synchronization, by employing subluminal incident ST wave packets. However, the blind synchronization scenario is more stringent because the necessary increase in group velocity with incident angle is larger: this change must counterbalance the increase in propagation distance within both media. As such, an incident ST wave packet with larger $\widetilde{n}_{1}$ (deeper within the subluminal regime) is required for blind synchronization compared to isochronous ST wave packets. Pursuing the concept of blind synchronization further requires increasing the propagation distance over which this effect is realized by reducing the so-called `spectral uncertainty' \cite{Yessenov19OE,Bhaduri19OL,Kondakci19OL}, which is the unavoidable `fuzziness' in the association between spatial and temporal frequencies in the wave packet spectrum. In general, reducing the spectral uncertainty to increase the propagation distance requires increasing the system's numerical aperture.

The work reported in this paper sequence, in addition to \cite{Bhaduri20NP}, has examined the refraction of ST wave packets in the simplest scenario: at a planar interface between two non-dispersive, homogeneous, isotropic dielectrics. Future work will be directed to an exploration of other important configurations. Paramount amongst these are the cases of dispersive and anisotropic media, which will be critical as a prelude to studying nonlinear interactions involving ST wave packets. Additionally, it will be useful to study the refraction of ST wave packets in which both transverse dimensions are included (ST needles rather than ST sheets), and ST wave packets in which a parameter is controlled along the propagation axis, such as wave packets that accelerate or decelerate \cite{Yessenov20PRL2}, or are endowed with axial spectral encoding \cite{Motz21PRA}.

In conclusion, we have verified experimentally the law of refraction for ST wave packets obliquely incident on a planar interface between two non-dispersive, homogeneous, isotropic dielectrics (Eq.~\ref{Eq:ObliqueIncidenceLaw}). This law was verified first by changing the group velocity of the incident ST wave packet at a fixed angle of incidence, and, second, by changing the angle of incidence while holding the group velocity of the incident wave packet fixed. Our measurements reveal that the group velocity of the transmitted wave packet \textit{increases} with the incident angle in the subluminal regime (when the index of the second medium is higher than that of the first); conversely, the group velocity of the transmitted wave packet \textit{decreases} with incident angle in the superluminal regime. Furthermore, we observed the shift that occurs in the conditions for group-velocity invariance and group-velocity inversion at oblique incidence with respect to those at normal incidence. These findings made possible the first observation of blind synchronization, whereby ST wave packets produced from a fixed source (Tx) are arranged to arrive simultaneously at different receivers (Rx) at \textit{a priori} unknown positions, except that they are located at a fixed depth beyond an interface.

\section*{Funding}
U.S. Office of Naval Research (ONR) contract N00014-17-1-2458.

\vspace{2mm}
\noindent
\textbf{Disclosures.} The authors declare no conflicts of interest.

\bibliography{diffraction}

\end{document}